\documentclass{article}
\usepackage{epsfig}
\newcommand{\be} {\begin{equation}}
\newcommand{\ee} {\end{equation}}
\newcommand{\bdm} {\begin{displaymath}}
\newcommand{\edm} {\end{displaymath}}
\newcommand{\bc} {\begin{center}}
\newcommand{\ec} {\end{center}}
\newcommand{\beqa} {\begin{eqnarray}}
\newcommand{\eeqa} {\end{eqnarray}}
\newcommand{\nn} {\nonumber}

\begin{document}

\begin{flushright}
M/C-TH 00/08\\
HD-THEP-00-50\\
October 2000\\
\end{flushright}
\begin{center}
\vspace*{1.5cm}

{\large \bf Diffractive Exclusive Photon Production in Deep Inelastic 
Scattering}

\vspace*{1cm}

A Donnachie\\
\medskip
Department of Physics and Astronomy, University of Manchester,\\
Manchester, M13 9PL, England\footnote{{\tt email: ad@theory.ph.man.ac.uk; 
h.g.dosch@thphys.uni-heidelberg.de}}\\
\vspace{0.5cm}
and\\ 
\vspace{0.5cm}
H G Dosch\\
\medskip
Institut f{\" u}r Theoretische Physik der Universit{\" a}t Heidelberg\\
Philosophenweg 16, 69120 Heidelberg, Germany\\
\medskip
and\\
\medskip
The Erwin Schr{\"o}dinger Institute for Mathematical Physics\\
Boltzmanngasse 9, A-1090 Wien, Austria 
\end{center}
\vspace*{1.6cm}

\begin{center}
{\bf Abstract}
\end{center}
\bigskip
Predictions for Deep Virtual Compton Scattering are obtained in a 
two component dipole model of diffraction. The model automatically includes 
hard and soft components and implicitly allows for ``hadronic'' contributions 
via large dipoles. It is also applicable to real Compton Scattering, which
provides an important constraint. 

\newpage

\section{Introduction}

Deep Virtual Compton Scattering (DVCS) on protons, $\gamma^* p \to \gamma p$,  
is an important reaction for the study of diffraction. In the standard 
perturbative QCD approach the amplitude is described by skewed parton  
distributions \cite{dvcs1} for $Q^2 > Q_0^2$, for some non-zero $Q_0^2$.
The skewed parton distributions correspond to operator products evaluated  
between protons of unequal momenta. These are generalisations of the familiar  
parton distributions of deep inelastic scattering, and like them satisfy 
perturbative evolution equations \cite{dvcs2} which enable them to be 
evaluated at all $Q^2$ in terms of an assumed input at some appropriate 
$Q^2 = Q_0^2$. 

Preliminary data have been presented \cite{H100} which are 
consistent with QCD predictions \cite{FFS98}.
The QCD calculations require the input skewed parton distributions at the
reference point $Q_0^2$. In \cite{FFS98} these are obtained by estimating 
their ratio to ``ordinary'' parton distributions at $Q_0^2 = 2.5$ GeV$^2$ 
using arguments based on the aligned jet model \cite{BK73}. In practice 
this is equivalent \cite{DGS00} to the simplest diagonal generalised vector 
meson dominance model \cite{SS72}. The r{\^ o}le of vector mesons, 
particularly the $\rho$, in providing a ``hadronic'' contribution to DVCS 
via the vector-meson-dominance mechanism $\gamma^* p \to V p$, $V \to \gamma$ 
has been calculated in \cite{DGS00} and shown to be important at values of 
$Q^2$ well beyond the input reference point $Q_0^2$. 

The model of $\rho$ electroproduction used in \cite{DGS00} is a two-component 
model, combining soft and hard contributions which are respectively primarily 
non-perturbative and perturbative in origin. There is considerable evidence 
in $\gamma^* p$ interactions that the nominally perturbative regime can be 
strongly influenced by non-perturbative effects. This is an obvious feature of 
dipole models of deep inelastic scattering \cite{NNPZ97}$-$\cite{FMS00}, 
where the contribution from large dipoles can extend to significantly large 
values of $Q^2$; and in two-component models \cite{DL98}$-$\cite{GLMN99}, 
in which there is a non-perturbative term at all $Q^2$ by construction.

Here we present the results of a model which incorporates these effects not in 
the language of parton distributions but rather that of dipole cross sections. 
In the framework of functional integration a two-component dipole cross 
section 
is constructed based on the approach of \cite{Rue99,DDR99}.  The model covers 
the complete range from the real $\gamma p$ cross section through DVCS,with 
one photon virtual, to deep inelastic scattering, with both photons virtual. 
The model as applied to DVCS contains no free parameters at $t = t_{\rm min}$: 
they have already been determined from the $p p$ and $\gamma^* p$ total cross
sections. To obtain the integrated cross section requires knowledge of the 
logarithmic slope of the differential cross section. This can also be 
determined from the model, and for $\rho$ electroproduction \cite{DGKP97} is 
in good agreement with the experimental value over the relevant range of 
$Q^2$. We use this as a reasonable estimate in the present dipole approach. 

\section{The model}

We present a calculation for the cross section of Compton scattering with
different virtualities of the photons. The treatment is based on a
functional approach of high energy scattering with small momentum transfer
\cite{Nac91} in which the scattering of a $q \bar q$ pair is expressed as the
functional average of a Wegner-Wilson loop with light-like sides 
\cite{KD91,DFK94}. The functional average is performed in a model for 
non-perturbative QCD \cite{Dos87,DS88}. The model depends essentially on 
two typically non-perturbative parameters: the strength of the gluon 
correlator, given through the gluon condensate $\langle g^2FF\rangle$, and 
its correlation length $a$. The linear confining potential is also defined 
in terms of these two parameters \cite{Dos87,DS88}. Nucleons are most simply 
treated as a quark-diquark system, that is effectively as a dipole, although 
this is not essential. The two parameters of the model were obtained by 
fitting the 
isoscalar part of $\bar{p}p$ and $p p$ scattering at $W = 20$ GeV, and were 
found to be $\langle g^2FF\rangle = 2.49$ GeV$^4$ and $a = 0.346$ fm. These 
values are within the range determined from lattice calculations, and give the 
correct value for the slope of the confining potential.

In this approach scattering amplitudes with small momentum transfer can also
be calculated, but in this note we concentrate on forward scattering. Then t
he approach  becomes a simple dipole model  and the Compton scattering 
amplitude can be expressed as an integral over the product of the dipole-proton 
cross  section $\sigma_d(R)$ and the overlap of the photon wave functions with 
virtualities $Q_1^2$ and $Q_2^2$ and helicity $\lambda$, $\rho^\lambda_\gamma
(Q_1^2,Q_2^2,R,z)$:
\begin{equation}
T(s,t=0) = 2 \pi \int_0^1 dz \int dR \,R\,
\rho^\lambda_\gamma(Q_1^2,Q_2^2,R,z)\,\sigma_d(R)  
\end{equation} 
where $z$ is the longitudinal momentum fraction of the quark in the photons
and $R$ is the radius of the quark-antiquark dipole, the weak dependence of 
the dipole cross section \cite{DGKP97} is neglected. The dipole cross section  
$\sigma_d(R)$ as evaluated in the model can be very well described by 
\begin{equation}
\sigma_d(R) = 0.098 \Big(\langle g^2 FF \rangle a^4\Big)^2 R
\Big(1-\exp\Big(-\frac{R}{3.1 a}\Big)\Big) 
\end{equation}
The dimensionless constant $\langle g^2 FF \rangle a^4$ has the numerical
value 23.8. If $a$ and $R$ are measured in fermi, the result is in millibarn.

As it stands the model has no energy dependence. The increase with energy of
hadronic cross sections as $W^{2\epsilon_{\rm soft}}$, where the intercept of
the pomeron trajectory is $1+\epsilon_{\rm soft}$ with $\epsilon_{\rm soft} 
\sim 0.09 \pm 0.01$, can be incorporated in two ways. Either one lets the 
radii of hadrons increase with $W^2$ \cite{DFK94},\cite{FP97}$-$\cite{PF00}, 
or one takes the model as a determination of the coefficient of pomeron 
exchange and includes a factor $(W/W_0)^{2\epsilon_{\rm soft}}$ with 
$W_0 = 20$ GeV. 
These two approaches give very similar results and we adopt the latter in 
this paper as it is the more convenient in the present context.

To incorporate the greater energy dependence observed in reactions with 
a hard scale, such as deep inelastic scattering, the two-pomeron approach of 
\cite{DL98} was adapted to the model in \cite{Rue99} and applied successfully 
to the photo- and electroproduction of vector mesons and to the proton 
structure function over a wide range of $Q^2$. As in \cite{DL98} it was found 
that the soft-pomeron contribution to the proton structure function initially 
increases with increasing $Q^2$, has a broad maximum in the region of 
5 GeV$^2$ and then decreases slowly as $Q^2$ decreases further. In the context 
of the present model this is a consequence of the decreasing interaction 
strength with decreasing dipole size.

It is worth recalling the salient features of this version of the two-pomeron
model to illustrate the distinction between the soft- and hard-pomeron 
contributions. Following  \cite{Rue99} it is assumed that all dipole 
amplitudes in which both dipoles are larger than the correlation length 
$a = 0.346$ fm are dominated by the soft pomeron, and the energy dependence 
is given by $(W/W_0)^{2\epsilon_{\rm soft}}$. This ensures that the hard 
pomeron has essentially no impact on purely hadronic scattering. If at least 
one of the dipoles is smaller than $a $ fm then the energy dependence 
is replaced by $(W/W_0)^{2\epsilon_{\rm hard}}$, with $\epsilon_{\rm hard}=
0.42$. This is the value found by \cite{DL98} and is larger than the value 
used in \cite{Rue99}. This difference in $\epsilon_{\rm hard}=0.42$ between 
the present paper and \cite{Rue99} is accounted for by a different 
treatment of the $z$ integration. In \cite{Rue99} there was a cutoff in the $z$
integration, requiring that $W\,z \,(1-z) > 0.2$ GeV in order to ensure that 
the quarks in the centre-of-mass frame have at least this energy. This $z$ 
cutoff 
introduces an additional energy dependence, particularly for highly virtual
photons. In our approach we follow the general lines of the dipole models and 
integrate $z$ from zero to one and therefore need a higher intercept for the 
hard component. It was found in \cite{Rue99} that the extrapolation of the 
non-perturbative model to high values of $Q^2$ overestimates the small dipole 
contributions and therefore the dipole cross section is put to zero below a 
radius $R_c=0.16$ fm.  

The quark-antiquark overlap density of photons with virtuality $Q_1^2$ and
$Q_2^2$ respectively and helicity $\lambda$ in lowest order perturbation
theory is
\beqa
\rho^0_\gamma(Q_1^2,Q_2^2,R,z)&=&{{2N_c\alpha}\over{\pi^2}}e_f^2Q_1\,Q_2\, 
z^2(1-z)^2
K_0(\epsilon_1 R)K_0(\epsilon_2 R)\nn\\
\rho^{\pm 1}_\gamma(,Q_1^2,Q_2^2,R,z)&=&{{2N_c\alpha}\over{\pi^2}}e_f^2
\Big((z^2+(1-z)^2)\epsilon_1 K_1(\epsilon_1 R)\epsilon_2 K_1(\epsilon_2
R)\nn\\
&&+m_f^2K_0(\epsilon_1 R)K_0(\epsilon_2 R)\Big). 
\label{photon_wf}
\eeqa
Here $e_f$ is the charge of the quark in units of the elementary charge; 
$m_f$ is the Lagrangian quark mass; $\epsilon_i=\sqrt{z(1-z)Q_i^2+m_f^2}$; 
$\lambda = 0$ indicates a longitudinal photon and
$\lambda = \pm 1$  indicates a transverse photon. The singularities of the
Bessel functions at  $R = 0$ do not cause any problems, since for the
evaluation of observable  amplitudes the density is multiplied by the dipole
cross section which is  proportional to $R^2$ for small values of R.

For longitudinal photons the factor $z^2(1-z)^2$ in the density $\rho^\lambda
_\gamma$ ensures that the main contribution comes from the region $z \approx 
{{1}\over{2}}$, and the relevant scale is ${{1}\over{4}}Q^2+m_f^2$. Thus 
contributions from large dipoles are suppressed at large $Q^2$. However for 
transverse photons the endpoints $z=0$ and $z=1$ are not suppressed and large 
dipoles can contribute to the perturbative regime even if the virtuality is 
quite high. The $z$ cutoff of \cite{Rue99} mentioned above is therefore most
important for large values of $Q_i^2$. 

The results (\ref{photon_wf}) are reliable for large values of $Q^2$ and/or  
large quark masses. For small values of $Q^2$ and light quarks the separation  
of the quark-antiquark pair can become large and confinement effects become  
important. An effective way in which to deal with this problem is to introduce 
a $Q^2$-dependent effective constituent quark mass  as an infrared regulator
in the photon wave function. This procedure is justified in \cite{DGP98} where
the value of the light quark effective mass at $Q^2=0$ is determined
from the two-point function of the vector current as $m_{0q}=0.21\pm 0.015$
GeV and for the strange quark as $m_{0s}\approx 0.31$ GeV. Since in this note 
we consider only processes where one photon is real we always use these 
effective mass values at $Q^2=0$. The charmed quark mass was taken as 
$m_c=1.3$ GeV. 

\section{Results}

\begin{figure}[t]
\leavevmode
\centering
\begin{minipage}{10cm}
\epsfxsize10cm
\epsffile{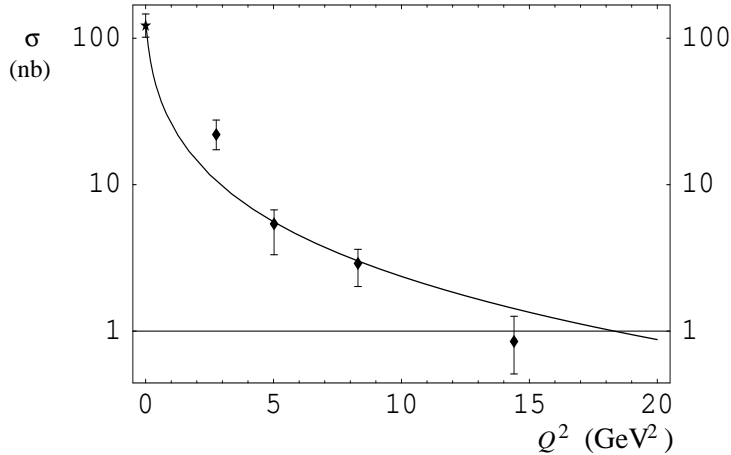}
\centering 
\end{minipage}
\caption{The integrated cross section for the reaction $\gamma^*\,p \to
\gamma\,p$ as function of the virtuality $Q^2$ of the incoming photon at an
average $\langle W \rangle= 75$ GeV} 
\label{DVCS_1}
\end{figure}

A convenient fit to the model calculation for $0\leq Q^2\leq 100$ GeV$^2$
(good in the amplitudes up to 1 \% for $Q^2 <20$ GeV$^2$ and 3 \% for higher $Q^2$ values) for
one photon real, the other one with virtuality $Q^2$  is given by:
\begin{eqnarray}
&&\frac{d \sigma}{dt}(Q^2,W,t=t_{min}) =\nn\\
&&\frac{1}{16 \pi} \Big(
r_{\rm soft}(Q^2)\big(W/W_0\big)^{0.16} 
 +r_{\rm hard}(Q^2)\big(W/W_0\big)^{0.84}~\Big)^2 311~\mu 
\rm{b/GeV}^2 
\end{eqnarray}
where $W_0 = 20$ GeV and
\begin{eqnarray}
r_{\rm soft}(Q^2) &=& \Big( 5.33 - 1.33 \exp(- 4 Q^2/Q_0^2)+5.37
\,Q^2/Q_0^2\Big)^{-1} \nn \\
r_{\rm hard}(Q^2) &=& \Big( 49.42 - 7.65 \exp(- 4 Q^2/Q_0^2)+ 4.94
\,Q^2/Q_0^2 \Big)^{-1}
\end{eqnarray}
with $Q_0^2 = 1$ GeV$^2$
The present calculation provides the $\gamma^* p \to \gamma p$ amplitude only
for forward scattering, that is at $t = t_{min} \approx-m_{proton}^2\, 
Q^2/s^2$, 
so it is necessary to make an assumption about the  $t$-dependence to obtain an
integrated cross section. As the process is dominated by light-quark dipoles,
it is reasonable to take the $t$-dependence to be similar to that for $\rho$
electroproduction. For definiteness we take the logarithmic slope to be $B =
7$ GeV$^{-2}$, which is the average value over the $Q^2$ range of the
preliminary H1 data \cite{H100}. 
Such a value of the average slope is also
the result of model calculation for electroproduction of $\rho$ mesons
\cite{DGKP97,DGS00b}.

\begin{figure}[t]
\leavevmode
\centering
\begin{minipage}{10cm}
\epsfxsize10cm
\epsffile{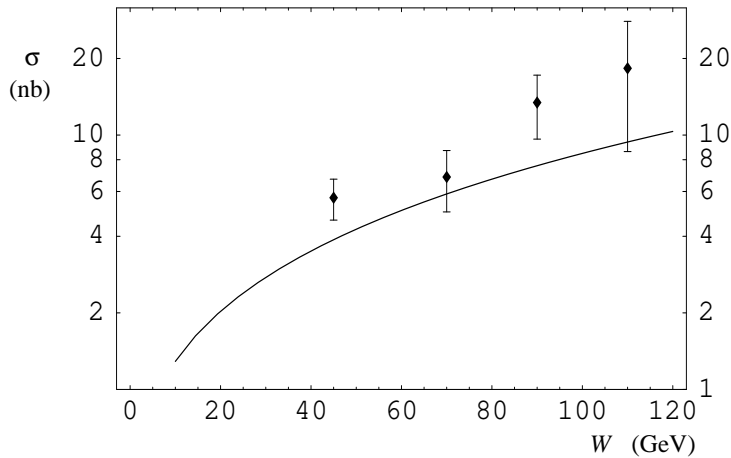}
\centering 
\end{minipage}
\caption{The integrated cross section for the reaction $\gamma^*\,p \to
\gamma\,p$ as function of the center-of-mass energy W at an average
virtuality of the incoming photon of $\langle Q^2\rangle = 4.5$ GeV$^2$}
\label{DVCS_2}
\end{figure}

The relation  between the $e p \to e p \gamma$ cross section and the 
$\gamma^* p
\to  \gamma p$ cross section is \cite{ZEUS98} \be {{d^2\sigma_{e p \to e p
\gamma}}\over{dW\,dQ^2}} = {{\alpha_{\rm em}}
\over{\pi}}{{W}\over{Q^2(W^2+Q^2-m_{\rm proton}^2)}}\big(1+(1-y)^2\big)
\sigma_{\gamma^* p \to \gamma p}, \label{gammap_to_ep}
\ee
with $y \equiv (W^2+Q^2-m_{\rm proton}^2)/(s-m_{\rm proton}^2)$. Here 
$\sqrt{s}$ 
is the centre-of-mass energy of the $ep$ system and $W$ is that of the 
$\gamma^* p$ system.

The $Q^2$ dependence and energy dependence of the predicted integrated 
$\gamma^* p \to \gamma p$ cross sections are given in figures \ref{DVCS_1}
and \ref{DVCS_2} respectively. The preliminary H1 data \cite{H100} are for 
$ep \to ep\gamma$ and contain a large background from the QED Bethe-Heitler 
process. However as the latter is purely real and as the QCD amplitude
is mainly imaginary (the real part is expected to be only about $10 - 15\%$ 
of the imaginary part), interference is not large and the Bethe-Heitler cross 
section can simply be subtracted. A comparison with the preliminary H1 data 
after subtraction of the Bethe-Heitler is made in figures \ref{DVCS_1} and
\ref{DVCS_2}, where we have converted the experimental $e p \to e p \gamma$ 
to $\gamma^* p \to \gamma p$, after subtraction of the Bethe-Heitler cross
section, using (\ref{gammap_to_ep}). The only serious discrepancy between the 
model and the preliminary data is at $Q^2 = 3.5$ GeV$^2$. This is reflected
in the normalisation of the integrated cross section in figure \ref{DVCS_2}
as the low-$Q^2$ point dominates this cross section. We simply note that the 
inclusion of the real $\gamma p$ data is seen to provide an important 
constraint on models.

\vfill\eject

\end{document}